\documentclass[aps,twocolumn,showpacs,preprintnumbers,amsmath,amssymb]{revtex4}

\usepackage{graphicx}
\usepackage{dcolumn}
\usepackage{bm}

\renewcommand{\vec}[1]{\bm{#1}}
\newcommand{\ket}[1]{\left |\mbox{$#1$}\right\rangle}
\newcommand{\bra}[1]{\left\langle\mbox{$#1$}\right |}

\newcommand{\hbarunit}{}
\begin{document}

\title{Effect of exchange interaction on fidelity of quantum state transfer from a photon qubit to an electron-spin qubit}
\author{Yoshiaki Rikitake and Hiroshi Imamura}
\affiliation{
CREST and Nanotechnology Research Institute, National Institute of Advanced Industrial Science and Technology, 1-1-1 Umezono, Tsukuba, Ibaraki 305-8568, Japan}


\begin{abstract}
 We analyzed the fidelity of the quantum state transfer (QST)
 from a photon-polarization qubit to an electron-spin-polarization qubit
 in a semiconductor quantum dot,
 with special attention to the exchange interaction
 between the electron and the simultaneously created hole. 
 In order to realize a high-fidelity QST 
 we had to separate the electron and hole as soon as possible, 
 since the electron-hole exchange interaction modifies 
 the orientation of the electron spin. 
 Thus, we propose a double-dot structure 
 to separate the electron and hole quickly, 
 and show that the fidelity of the QST can reach
 as high as 0.996 if the resonant tunneling condition is satisfied.
\end{abstract}

\maketitle

Quantum state transfer (QST) has attracted enormous attention 
as one of the key concepts in quantum information science
\cite{vrijen01,duan01,kosaka03,muto05,ament06}.
Quantum information can take several different forms 
such as photons, nuclear spin of atoms, 
and electron spin of quantum dots. 
All of these physical realizations of quantum information 
are called  ``qubits''.
Since each qubit has its own merits and demerits, 
we have to choose the right qubit for each process. 
The photon-polarization qubit is the most convenient medium 
for sharing quantum information between distant locations\cite{gisin02}.
Presently, we can distribute quantum keys over 122km 
of standard telecom fiber\cite{gobby04}. 
On the other hand, the electron-spin qubit is 
the most convenient medium for quantum gate and quantum memory 
in a semiconductor quantum dot\cite{loss98,taylor03,petta05,petta05a}
since coupling among the qubits can easily be controlled 
by gate voltage. 
Electron-spin qubits are a promising candidate 
for the realization of a scalable quantum computer. 
It is then a logical next step 
to study the QST from a photon qubit to an electron-spin qubit 
in order to construct efficient quantum information processing devices. 

In 2001, Vrijen and Yablonovitch proposed a spin-coherent 
semiconductor photo-detector which transfers the quantum information 
from a photon-polarization qubit to an electron-spin qubit. 
Such a quantum-state-coherent photo-detector is 
a basic element of a quantum repeater
\cite{briegel98,childress05,taylor05},
which enables us to drastically expand 
the distance of quantum key distribution.
They showed that the well-known optical orientation 
in semiconductor heterostructure can be used 
for the QST.

\begin{figure}[t]
 \begin{tabular}{lll}
 (a) & & (b)
 \\
 \includegraphics[width=0.4\columnwidth]{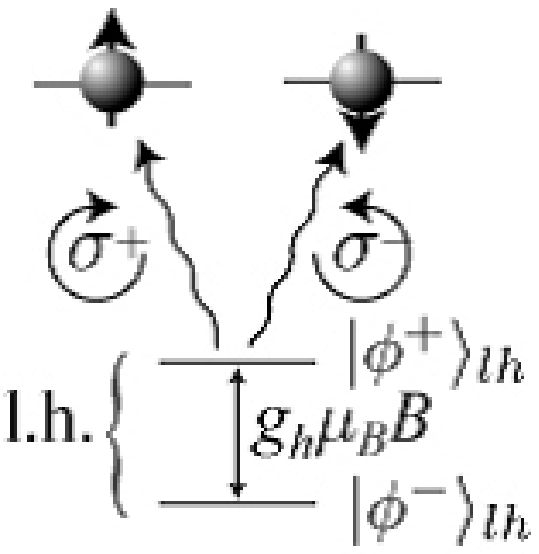}
 & \qquad &
 \includegraphics[width=0.4\columnwidth]{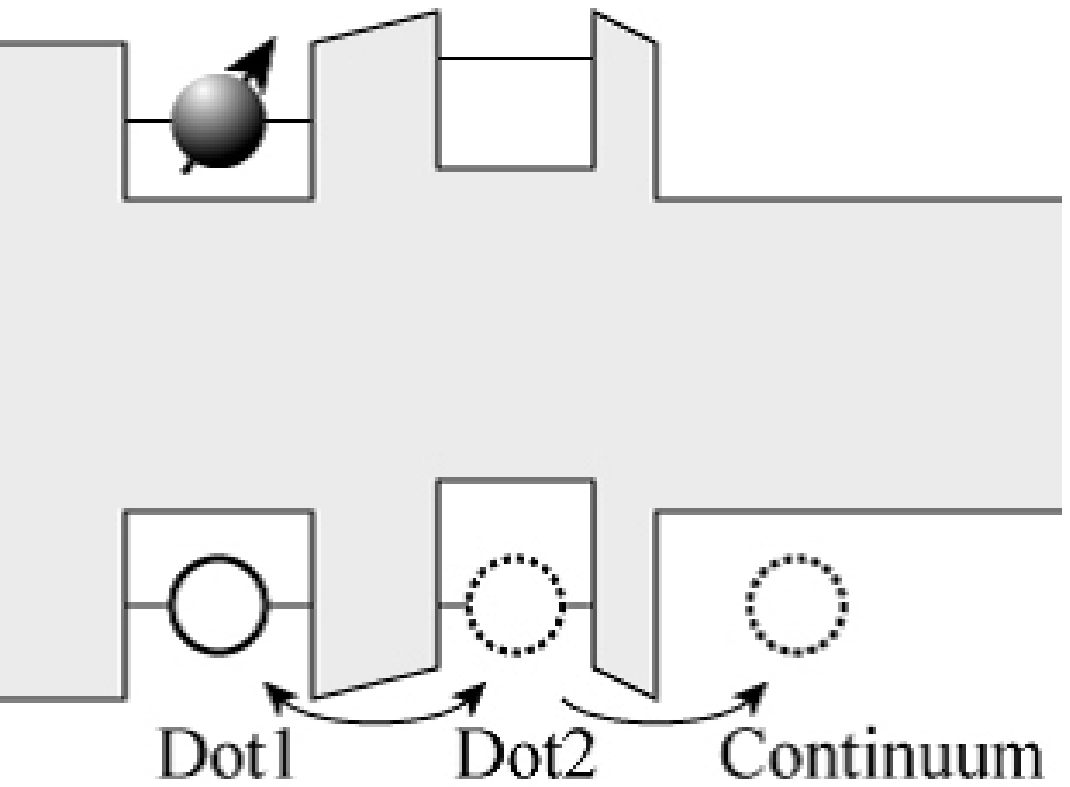}
 \end{tabular}
 \caption{(a) Selection rule for the quantum-state transfer from photon polarization to electron spin. The light-hole levels are split into $\ket{\psi^{+}}_{lh}$ and $\ket{\psi^{-}}_{lh}$ by the applied magnetic field $B$. From the $\ket{\psi^{+}}_{lh}$ state, the electron with $\ket{\uparrow}$ ($\ket{\downarrow}$) spin state is optically excited by the right-handed (left-handed) circularly polarized photon $\ket{\sigma^{+}}$ ($\ket{\sigma^{-}}$). 
 (b) Energy band of the system. The electron-hole pair is excited in dot1. The created hole is extracted from dot1 to the continuum of the hole via dot2.}
 \label{fig:Model}
\end{figure}

The spin-coherent semiconductor photo-detector has 
an optically active quantum well 
where the quantum information is transferred 
from photon polarization to electron spin. 
The $\vec{k}$-vector of the incident photon is parallel 
to the growth direction of the well. 
The energy levels of the well are shown in Fig. \ref{fig:Model} (a).
In order to carry out the photon-spin QST, the spin states
$\ket{\uparrow}$ and $\ket{\downarrow}$ of an electron should be
degenerate in the presence of a magnetic field. 
Therefore, we have to tune the electron spin $g$-factor 
to be zero, $g_e=0$, 
with the help of $g$-factor engineering
\cite{kiselev98,ivchenko98,matveev00,kosaka01,salis01,salis03,nitta03,lin04,nitta04}.
The $g$-factor engineering can be realized by using the proximity of
the electron wave function into the barrier layer.
In the quantum well system, 
the $g_e$ can be estimated as $g_e=w g_W+(1-w)g_B$,
where $g_W$ and $g_B$ are the $g$-factor of the well and
that of the barrier, respectively, and $w$ is the occupation
probability of the electron in the well.
Appropriate choice of the structure and the material enables us to
obtain $g_e=0$.
The degeneracy between the heavy-hole states and light-hole states
is lifted if the material is placed under tensile strain
\cite{lin91,nakaoka04,nakaoka05}.
The uniform magnetic field $B$ is applied along the $x$-direction 
to lift the degeneracy of the light-hole states 
$\ket{\psi^{+}}_{lh}$$=(\ket{J\! = \! 3/2,m_J\! = \! 1/2}$$+\ket{J\! = \! 3/2,m_J\! = \! -1/2})/\sqrt{2}$ 
and $\ket{\psi^{-}}_{lh}$$=(\ket{J\! = \! 3/2,m_J\! = \! 1/2}$$-\ket{J\! = \! 3/2,m_J\! = \! -1/2})/\sqrt{2}$.
The Zeeman splitting between these two states is 
given by $g_{h}\mu_B B$, 
where $\mu_B$ is the Bohr magneton. 
The Zeeman splitting of the electron spin states is assumed to be zero. 
According to the selection rule, 
the electron with the spin up state along the $z$-direction 
$\ket{\uparrow}$ is excited in the quantum dot by a right-handed
circularly polarized photon $\ket{\sigma^{+}}$. 
Similarly, a left-handed circularly polarized photon 
$\ket{\sigma^{-}}$ excites the electron in the spin-down state 
$\ket{\downarrow}$. 
In these two cases, a hole in the $\ket{\psi^{+}}_{lh}$ state is 
created in the dot simultaneously. 
After elimination of the hole, the superposition of the polarized photon
$\alpha_{+}\ket{\sigma^{+}}+\alpha_{-}\ket{\sigma^{-}}$ 
is transferred to the superposition of the electron spin 
$\alpha_{+}\ket{\uparrow}+\alpha_{-}\ket{\downarrow}$.

One of the main obstacles to high-fidelity QST in
a spin-coherent semiconductor photo detector is 
the exchange interaction between the electron 
and the simultaneously created hole. 
In this paper, 
we analyze the effect of the exchange interaction 
on the fidelity of the QST from a photon-polarization qubit 
to an electron-spin qubit. 
For high-fidelity QST we have to extract the hole 
as soon as possible. 
We propose a double-well structure 
to separate the electron and hole quickly via resonant tunneling.
Quick extraction of the carrier using resonant tunneling 
in the double-well structure was extensively studied 
by Gurvitz\cite{gurvitz91}, and experimentally demonstrated 
by Cohen\cite{cohen93}. 
Using the double-well structure, 
quick extraction of the hole can be realized 
without thinning the barrier width,
and then we can minimizes the deviation of $g_e$ from zero.
We solved the Schr{\" o}dinger equations of the photo-detector 
using realistic parameters of semiconductor heterostructure 
and showed that the fidelity of the QST can reach 
as high as 0.996 under the resonant tunneling condition.

The system we consider is the semiconductor heterostructure 
shown in Fig.\ref{fig:Model}(b), which consists of two dots, 
dot1 and dot2. 
Dot2 is connected with the continuum through the tunneling barrier. 
The strong confinement of the dot structure prolongs 
the electron-spin coherence time $T_2$. 
The gates are attached to the dots to control the energy levels. 
Since our interest is in the effect of the electron-hole exchange
interaction 
on the fidelity of the QST, 
we restricted our study to the dynamics of the system 
after the electron-hole pair is created in dot1 
by an incident photon. 
We assume that the dipolar interaction is so small 
that we can neglect the recombination process. 
We also assume that the mismatch of the electron energy levels 
between two quantum dots is much larger than 
the inter-dot tunneling coupling, and therefore 
that the electron is localized in dot1.

The wave function of the system can be written as 
\begin{align}
 \ket{\Psi(t)} 
 & = 
 \sum_{s=\uparrow\downarrow}\phi_{1s}(t)\ket{sh_1}
 +
 \sum_{s=\uparrow\downarrow}\phi_{2s}(t)\ket{sh_2}
 \nonumber
 \\
 & +
 \sum_{s=\uparrow\downarrow}\sum_{l}\psi_{ls}(t)\ket{sl},
 \label{eq:Psi}
\end{align}
where $s=\uparrow,\downarrow$ denotes the electron state with spin $s$, 
$h_{1(2)}$ the hole state in dot1(2), 
and $l$ the hole state in the continuum.
Note that the state of the hole in the dot1 $h_{1}$
is assumed to be restricted into the top-most light-hole state 
$\ket{\psi^{+}}$.
Here, $\phi_{1s}(t)$, $\phi_{2s}(t)$, 
and $\psi_{ls}(t)$ are coefficients 
to be determined by solving the Schr\"{o}dinger equation.

The Hamiltonian of the system is expressed as
\begin{align}
 H 
 & =
 \sum_{s=\uparrow\downarrow} \hbarunit(\omega_e+\omega_1)
 \ket{sh_1}\!\bra{sh_1}
 +
 \sum_{s=\uparrow\downarrow} \hbarunit\omega_J
 \ket{sh_1}\!\bra{\bar{s}h_1}
 \nonumber 
 \\
 & +
 \sum_{s=\uparrow\downarrow}
 \{\hbarunit\delta\ket{sh_1}\!\bra{sh_2} + \text{h.c.}\}
  +
 \sum_{s=\uparrow\downarrow}
 \hbarunit(\omega_e+\omega_2)\ket{sh_2}\!\bra{sh_2}
 \nonumber
 \\
 & +
 \sum_{s=\uparrow\downarrow}\sum_{l}
 \{\hbarunit W_l \ket{sh_2}\!\bra{sl} + \text{h.c.}\}
 \nonumber
 \\
 & + 
 \sum_{s=\uparrow\downarrow}\sum_{l}
 \hbarunit(\omega_e+\omega_l)\ket{sl}\!\bra{sl},
 \label{eq:Hamiltonian}
\end{align}
where $\hbarunit\omega_e$ is the energy level of the electron in dot1, 
$\hbarunit\omega_{1(2)}$ the hole energy level in dot1(2), 
$\hbarunit\omega_J$ the electron-hole exchange interaction, 
$\hbarunit\omega_l$ the hole energy level in the continuum, 
$\delta$ the coupling between dot1 and dot2, 
$W_l$ the coupling between dot2 and continuum state $l$, 
and $\bar{s}$ the electron spin opposite to $s$.
Here, we set $\hbar=1$.
In a zincblende crystal, the electron-hole exchange interaction
is given by $ a\, \vec{s}\cdot\vec{J} + 
b\,\sum_{\lambda=x,y,z} s_{\lambda}J_{\lambda}^3$
\cite{ivchenkobook},
where $a, b$ are coefficients, $\vec{s}$ and $\vec{J}$
represent the electron and hole spin, respectively.
In Eq. (\ref{eq:Hamiltonian}),
we consider the coupling term between two degenerated states
$\ket{\uparrow h_1}$ and $\ket{\downarrow h_1}$,
$\omega_J=(2a+5b)/4$.
We can neglect the coupling with the other states since the 
typical energy scale of the exchange interaction ($\sim 10\mu\text{eV}$) is
much smaller than the Zeeman splitting between the light hole
states ( $\sim$ 1 meV at $B=5\text{T}$).


The dynamics of the system are obtained 
by solving the following Schr\"{o}dinger equation:
\begin{align}
 \dot{\phi}_{1s}(t) 
 & =
 - i (\omega_e + \omega_1) \phi_{1s}(t) - i\omega_J \phi_{1\bar{s}}(t)
 - i \delta \phi_{2s}(t),
 \label{eq:dot_phi_1s}
 \\
 \dot{\phi}_{2s}(t) 
 & =
 - i (\omega_e + \omega_2) \phi_{2s}(t) - i \delta^{\ast} \phi_{1s}(t)
 \nonumber
 \\
 &\quad
 - i \sum_l W_l\psi_{ls}(t),
 \label{eq:dot_phi_2s}
 \\
 \dot{\psi}_{ls}(t)
& =
 - i (\omega_e + \omega_l) \psi_{ls}(t) - i W_l^{\ast}\phi_{2s}.
 \label{eq:dot_psi_ls}
\end{align}
These equations can be simplified by changing the electron spin basis 
from the eigenstates ($\ket{\uparrow},\ket{\downarrow}$) of $\sigma_z$ 
to the eigenstates ($\ket{+}$,$\ket{-}$) of $\sigma_x$, introducing
\begin{align}
 \phi_{1\pm}(t) 
 & = 
 (\phi_{1\uparrow}(t)\pm\phi_{1\downarrow}(t))/\sqrt{2},
 \\
 \phi_{2\pm}(t) 
 & = 
 (\phi_{2\uparrow}(t)\pm\phi_{2\downarrow}(t))/\sqrt{2},
 \\
 \psi_{l\pm}(t)
 & =
 (\psi_{l\uparrow}(t)\pm\psi_{l\downarrow}(t))/\sqrt{2}.
\end{align}
Eqs (\ref{eq:dot_phi_1s})-(\ref{eq:dot_psi_ls}) are rewritten as
\begin{align}
 \dot{\phi}_{1\sigma}(t) 
 & =
 - i (\omega_e + \omega_1 + \omega_{J\sigma}) \phi_{1\sigma}(t)
 - i \delta \phi_{2\sigma}(t),
 \label{eq:dot_phi_1sigma}
 \\
 \dot{\phi}_{2\sigma}(t) 
 & =
 - i (\omega_e + \omega_2) \phi_{2\sigma}(t) 
 - i \delta^{\ast} \phi_{1\sigma}(t)
 \nonumber
 \\
 & \quad - i \sum_l W_l\psi_{l\sigma}(t),
 \label{eq:dot_phi_2sigma}
 \\
 \dot{\psi}_{l\sigma}(t)
 & =
 - i (\omega_e + \omega_l) \psi_{l\sigma}(t) - i W_l^{\ast}\phi_{2\sigma}(t),
 \label{eq:dot_psi_lsigma}
\end{align}
where $\sigma=\pm$, $\omega_{J\pm} = \pm \omega_J$. 
One can easily see that Eqs 
(\ref{eq:dot_phi_1sigma})-(\ref{eq:dot_psi_lsigma}) are separable 
with respect to the index $\sigma=\pm$. 
The solution of Eq.(\ref{eq:dot_psi_lsigma}) is obtained as
\begin{align}
 \psi_{l\sigma}(t)
 & =
 -i W_l^{\ast}\int_0^{t} dt^{\prime} e^{-i(\omega_e+\omega_l)(t-t^{\prime})}
 \phi_{2\sigma}(t^{\prime}).
 \label{eq:psi_lsigma}
\end{align}
The tunneling process of the hole from the dot2 to the continuum
is characterized by the spectral density function
$\gamma_h(\omega)\equiv\pi\sum_{l}|W_l|^2 \delta(\omega-\omega_l)$.
Given the density of the state of the hole in the continuum is
dense around $\omega\sim\omega_2$,
we can treat $\gamma_h(\omega)$ as a constant,
which corresponds to the Markov approximation.
Substituting Eq.(\ref{eq:psi_lsigma}) into Eq.(\ref{eq:dot_phi_2sigma}), 
and applying the Markov approximation,
we have
\begin{align}
 \dot{\phi}_{2\sigma}(t) 
 & =
 - i (\omega_e + \omega_2 - i \gamma_h) \phi_{2\sigma}(t) 
 - i \delta^{\ast} \psi_{1\sigma}(t).
 \label{eq:dot_phi_2sigma_new}
\end{align}
Here, $\gamma_h$ represents the tunneling rate of the hole
from the dot2 to the continuum.

Applying the Laplace transformation
$\hat{\phi}_{i\sigma}(p) = \int_0^{\infty} dt e^{-pt}\phi_{i\sigma}(t)$,
Eqs (\ref{eq:dot_phi_1sigma}) and (\ref{eq:dot_phi_2sigma_new}) 
can be expressed as
\begin{align}
&\!\! p\hat{\phi}_{1\sigma}(p) \! - \! \beta_{\sigma}
  =
 \! - \! i( \omega_e+\omega_1+\omega_{J\sigma})\hat{\phi}_{1\sigma}(p)
 \! - \! i \delta \hat{\phi}_{2\sigma}(p),
 \label{eq:p_hat_phi_1sigma}
 \\
&\!\! p\hat{\phi}_{2\sigma}(p)
 =
 - i( \omega_e+\omega_2-i\gamma_h)\hat{\phi}_{2\sigma}(p)
 - i \delta^{\ast} \hat{\phi}_{1\sigma}(p),
 \label{eq:p_hat_phi_2sigma}
\end{align}
where $\beta_{\sigma}$ are the coefficients 
for the linear combination of electron spin states $\ket{\pm}$ at
$t=0$. 
Then we obtain
\begin{align}
 \hat{\phi}_{1\sigma}(p) 
 & = 
 \beta_{\sigma}
 \Big[
 p + i(\omega_e+\omega_1+\omega_{J\sigma})
 \nonumber
 \\
 &\qquad
 + \dfrac{|\delta|^2}{p+i(\omega_e+\omega_2-i\gamma_h)}
 \Big],
 \label{eq:hat_phi_1sigma}
 \\
 \hat{\phi}_{2\sigma}(p) 
 & = 
 \dfrac{-i\delta^{\ast}}{p+i(\omega_e+\omega_2-i\gamma_h)}
 \hat{\phi}_{1\sigma}(p).
 \label{eq:hat_phi_2sigma}
\end{align}

The fidelity of the QST is defined as
$
 F =
 \bra{\Psi(0)}\rho(\infty)\ket{\Psi(0)},
 \label{eq:F_dif}
$
where $\rho(t)$ is the reduced density matrix, 
and $\ket{\Psi(0)}=\beta_{+}\ket{+}+\beta_{-}\ket{-}$ 
the initial state of the spin. 
Each component of $\rho(t)$ is defined as
\begin{align}
 \rho_{\sigma\sigma^{\prime}}(t)
 & =
 \phi_{1\sigma}(t) \phi_{1\sigma^{\prime}}^{\ast}(t)
 +
 \phi_{2\sigma}(t) \phi_{2\sigma^{\prime}}^{\ast}(t)
 \nonumber
 \\
 &\quad
 +
 \sum_l \psi_{l\sigma}(t) \psi_{l\sigma^{\prime}}^{\ast}(t).
\end{align}
In the limit of $t\to\infty$, 
the hole is in the continuum and 
$\phi_{1\sigma}(\infty)=\phi_{2\sigma}(\infty)=0$. Hence, we have 
$\rho_{\sigma\sigma^{\prime}}(\infty) = \sum_l {\psi}_{l\sigma}(\infty) 
{\psi}_{l\sigma^{\prime}}^{\ast}(\infty)$.
The reduced density matrix $\rho_{\sigma\sigma^{\prime}}(\infty)$ can 
be easily evaluated by moving to the interaction picture. 
In the interaction picture, 
the reduced density matrix is expressed as 
$\rho_{\sigma\sigma^{\prime}}(\infty) = \sum_l \tilde{\psi}_{l\sigma}(\infty) 
\tilde{\psi}_{l\sigma^{\prime}}^{\ast}(\infty)$, 
where 
$\tilde{\psi}_{l\sigma}(t)\equiv e^{i(\omega_e+\omega_l)t}\psi_{l\sigma}(t)$.
From Eqs (\ref{eq:psi_lsigma}) and (\ref{eq:hat_phi_2sigma}),
$\tilde{\psi}_{l\sigma}(t)$ is given by
\begin{align}
 \tilde{\psi}_{l\sigma}(\infty) 
 & =
 -iW_{l}^{\ast}\hat{\phi}_{2\sigma}(-i(\omega_e+\omega_l))
 =
 \beta_{\sigma}
 \dfrac{\delta W_l}{f_{\sigma}(\omega_l)},
 \label{eq:tilde_psi_lsigma_infty}
\end{align}
where
$
 f_{\sigma}(\omega)
  =
 (\omega_1-\omega + \omega_{J\sigma})
 (\omega_2-\omega - i \gamma_h) 
 - |\delta|^2.
$
Thus we have
$
 \rho_{\sigma\sigma^{\prime}}(\infty)
  =
 \beta_{\sigma}\beta_{\sigma^{\prime}}^{\ast} I_{\sigma\sigma^{\prime}},
$
where
\begin{align}
 I_{\sigma\sigma^{\prime}}
 & =
 \dfrac{|\delta|^2 \gamma_h}{\pi}
 \int d\omega 
 \dfrac{1}{f_{\sigma}(\omega) f_{\sigma^{\prime}}^{\ast}(\omega)}.
 \label{eq:I}
\end{align}
Finally, the fidelity of the QST is obtained as
\begin{align}
 F
 & =
 1 - 2|\beta_{+}|^2|\beta_{-}|^2(1-\Re I_{+-}).
 \label{eq:F}
\end{align}

\begin{figure}[t]
 \includegraphics[width=0.7\columnwidth]{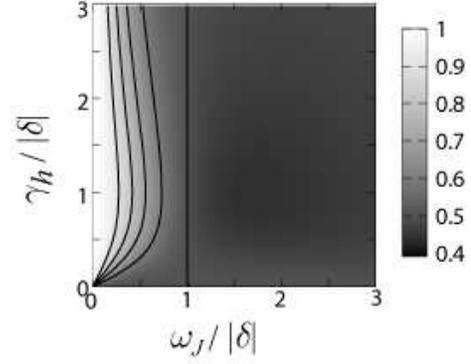}
 \caption{Contour plot of the fidelity $F$ as a function of $\omega_{J}/|\delta|$ and $\gamma_h/|\delta|$. The initial spin state is assumed to be $\ket{\uparrow}$.}
 \label{fig:F}
\end{figure}

Equation (\ref{eq:F}) shows that the fidelity depends 
strongly on the initial state of the electron spin, 
$\beta_{\sigma}$. 
If the initial state of the electron spin is
$(\ket{\uparrow}+\ket{\downarrow})/\sqrt{2}$ ($\beta_+=1,\beta_-=0$) 
or $(\ket{\uparrow}-\ket{\downarrow})/\sqrt{2}$ ($\beta_+=0,\beta_-=1$), 
the electron-hole exchange interaction does not affect the fidelity 
since the initial state is the eigenstate of the electron-hole exchange 
interaction. 
For the general initial states with $|\beta_+|^2|\beta_-|^2\neq 0$, 
the fidelity is reduced from unity by the electron-hole exchange interaction.
In Fig. \ref{fig:F}, 
we plot the fidelity $F$ as a function of 
the electron-hole exchange interaction, $\omega_J$, 
and the tunneling rate of the hole from dot2 to the continuum,
$\gamma_h$. 
We assume that the hole energy levels in dot1 and dot2 are the same, 
i.e., $\omega_1=\omega_2$. 
The initial state of the electron spin is taken to be 
$\ket{\uparrow}$ ($|\beta_+|^2=|\beta_-|^2=1/2$). 
These two axis values are normalized by the inter-dot coupling $|\delta|$.

As shown in Fig. \ref{fig:F}, 
the fidelity $F$ is a monotonic decreasing function of $\omega_J$ 
for $\omega_J<|\delta|$. 
The fidelity becomes lower than $1/2$ for $\omega_J>|\delta|$ 
because the electron-hole exchange interaction flips 
the electron-spin state 
before the hole is extracted from dot1. 
Therefore, the first condition for high-fidelity QST is 
$\omega_J\ll |\delta|$. 
The second condition for high-fidelity QST is for $\gamma_h$. 
The fidelity is not a monotonic function of $\gamma_h$ but 
has a maximum around $\gamma_h\sim|\delta|$ 
as shown in Fig. \ref{fig:F}. 
For $\gamma_h\ll |\delta|$, the escape time of the hole is 
dominated by the tunneling rate from dot2 to the
continuum, $\gamma_h$. 
As we increase $\gamma_h$, the escape time of the hole decreases. 
Therefore, the fidelity increases with increasing $\gamma_h$ 
as long as $\gamma_h\ll|\delta|$. 
On the contrary, for $\gamma_h\gg |\delta|$, 
coherent oscillation between hole states in dot1 and dot2 
is suppressed by the strong coupling between dot2 and the continuum. 
Therefore, the hole tends to localize in dot1 and 
the fidelity decreases with increasing $\gamma_h$. 
The quickest extraction of the hole is performed at 
$\gamma_h \sim |\delta|$, which is called 
the resonant tunneling regime\cite{gurvitz91}.
Hence, the conditions for high-fidelity QST are given by 
$
 \omega_J\ll|\delta|\sim\gamma_h.
$

\begin{figure}[t]
 \includegraphics[width=0.7\columnwidth]
 {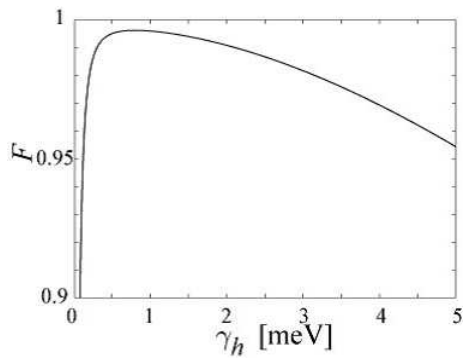}
 \caption{The fidelity $F$ is plotted as a function of the tunneling rate of the hole $\gamma_h$.}
 \label{fig:F_vs_gammah}
\end{figure}

We now proceed to the estimation of the fidelity of the electron spin 
using realistic parameters of GaAs/Al$_{0.8}$In$_{0.2}$As 
heterostructure\cite{nakayama92}. 
We can set the electron $g$-factor in the quantum dot to be zero 
by adjusting the thickness of the GaAs layer. 
The energy levels shown in Fig. \ref{fig:Model} (a) can be realized 
in this heterostructure since tensile strain is applied to the GaAs
layer.
The inter-dot coupling, $\delta$, can be calculated 
by considering the boundary conditions for the wave function 
for the hole\cite{gurvitz91,bardeen61,harrison61}. 
We estimate $|\delta|=0.8\text{meV}$ for GaAs/Al$_{0.8}$In$_{0.2}$As 
heterostructure with $g_e=0$. 
The typical value of the electron-hole exchange interaction 
in the quantum dot is 
$\hbarunit\omega_J=40\mu\text{eV}$\cite{gammon96,takagahara00,bayer02}. 
In Fig. \ref{fig:F_vs_gammah}, 
we plot the fidelity $F$ as a function of $\gamma_h$. 
As mentioned before, the fidelity takes its maximum value $F=0.996$ 
under the resonant condition at $\gamma_h=0.8\text{meV}$.

In conclusion, 
we analyzed the effect of the electron-hole exchange interaction 
on the QST in a spin-coherent semiconductor photo-detector. 
We have shown that the fidelity decreases as the strength 
of the exchange interaction increases, and that it depends on 
the initial state of the electron spin. 
We have also shown that a high-fidelity ($F\sim 0.996$) 
QST is possible using the double-dot structure 
under the resonant tunneling condition of a hole.

We acknowledge the valuable discussions 
we had with H. Kosaka and T. Takagahara.
This work was supported by CREST, MEXT.KAKENHI (No. 16710061), 
the NAREGI Nanoscience Project, and a NEDO Grant.

\bibliographystyle{apsrev}

\end{document}